\documentclass[journal]{IEEEtran}
\usepackage{cite}
\usepackage{amsmath,amssymb,amsfonts}
\usepackage{algorithmic}
\usepackage{graphicx}
\usepackage{textcomp}
\usepackage{xcolor}
\usepackage{physics}
\usepackage{array}
\usepackage[caption=false,font=normalsize,labelfont=sf,textfont=sf]{subfig}
\usepackage{fixltx2e}
\usepackage{stfloats}
\usepackage{url}
\usepackage{authblk}

\begin{document}

\title{Meta-Learning for GPU-Accelerated Quantum Many-Body Problems}

\author[1]{Yun-Hsuan Chen}
\author[2]{Jen-Yu Chang}
\author[3]{Tsung-Wei Huang}
\author[3,*]{En-Jui Kuo}

\affil[1]{Department of Intelligent Computing and Big Data, Chung Yuan Christian University, Taoyuan, Taiwan}
\affil[2]{Arete Honors Program, National Yang Ming Chiao Tung University, Hsinchu, Taiwan}
\affil[3]{Department of Electrophysics, National Yang Ming Chiao Tung University, Hsinchu, Taiwan}
\affil[*]{Corresponding author: \texttt{kuoenjui@nycu.edu.tw}}


\maketitle

\begin{abstract}
We explore the industrial and scientific applicability of the VQE–LSTM framework by integrating meta-learning with GPU accelerated quantum simulation using NVIDIA’s CUDA-Q (CUDAQ) platform. This work demonstrates how an LSTM–FC meta-initialization module can extend the practical reach of the Variational Quantum Eigensolver (VQE) in both chemistry and physics domains. In the chemical regime, the framework predicts ground-state energies of molecular Hamiltonians derived from \texttt{PySCF}, achieving near FCI accuracy while maintaining favorable $O(N^2)$ scaling with molecular size. In the physical counterpart, we applied the same model to quantized Simple Harmonic Motion systems (SHM), successfully reproducing its ground and excited states through VQE and Variational Quantum Deflation (VQD) methods. Benchmark results on NVIDIA GPUs reveal significant speedups over CPU-based implementations, validating CUDAQ’s capability to handle large-scale variational workloads efficiently. Overall, this study establishes VQE–LSTM as a viable and scalable approach for GPU accelerated quantum simulation, bridging quantum chemistry and condensed-matter physics through a unified, meta-learned initialization strategy.
\end{abstract}


\section{Introduction}
\IEEEPARstart{Q}{uantum} computing has emerged as a powerful paradigm for simulating strongly correlated systems that are intractable for classical computation.
By exploiting quantum superposition and entanglement, quantum processors can represent exponentially large Hilbert spaces using a polynomial number of qubits, enabling applications in quantum chemistry, condensed-matter physics, and many-body systems~\cite{preskill2018nisq,cao2019chemreview}.

Among near-term hybrid quantum--classical algorithms, the Variational Quantum Eigensolver (VQE) has become a dominant approach for estimating ground-state energies of parameterized Hamiltonians~\cite{peruzzo2014vqe,preskill2018nisq,cao2019chemreview,mcclean2020openfermion}.
VQE employs a parameterized quantum circuit (ansatz) whose expectation value with respect to a target Hamiltonian is minimized using a classical optimizer.
This hybrid structure makes VQE particularly well suited for Noisy Intermediate-Scale Quantum (NISQ) devices and large-scale quantum simulation workflows~\cite{preskill2018nisq}.

Despite its broad applicability, VQE faces significant optimization challenges. Random or heuristic parameter initialization frequently leads to slow convergence or stagnation in suboptimal local minima.
The variational energy landscape is highly non-convex, often exhibiting barren plateaus where gradients vanish exponentially with system size and depend sensitively on circuit structure and cost-function design~\cite{mcclean2020openfermion,cerezo2021costfunction}.
As system size increases, circuit evaluations grow rapidly and both gradient-free and gradient-based optimizers become increasingly sensitive to noise and initialization~\cite{fedorov2022survey}.
Extensions such as Variational Quantum Deflation (VQD) further compound the optimization difficulty when targeting excited states~\cite{harwood2022adiabaticvqe,higgott2019vqd}.

Recent progress in meta-learning provides a promising strategy for mitigating these challenges.
Meta-learning methods, particularly recurrent neural networks such as Long Short-Term Memory (LSTM), are capable of learning transferable optimization priors from previous tasks~\cite{hochreiter1997lstm}.
When applied to variational quantum algorithms, this approach enables neural networks to predict informed initial parameters by learning patterns across related Hamiltonians and optimization trajectories, substantially reducing quantum evaluations and improving robustness in practical workflows~\cite{cervera2021metavqe,verdon2019metaqnn,liu2022layervqe,chen2025quantumrainbow,chen2025framegeneration, lin2025metaqopt}.

In quantum chemistry, parameter initialization is especially critical due to the combinatorial growth of the Hilbert space, and prior studies have shown that neural-network-based strategies can improve VQE convergence for molecular and related physical systems~\cite{chang2025lstmfcvqe,tsai2020lstmmd}.
However, most existing Meta-VQE approaches are constrained to fixed system sizes due to limited computational scalability, preventing efficient training and evaluation across varying ansatz dimensions and qubit counts.

In this work, we investigate the applicability and scalability of an LSTM-FC-VQE framework integrated with NVIDIA’s CUDA-Q (CUDAQ) GPU accelerated quantum simulation platform~\cite{nvidia2024cudaq}.
Rather than proposing a new variational algorithm, we focus on evaluating how meta-learned initialization combined with large-scale GPU acceleration impacts convergence behavior, runtime efficiency, and scalability in realistic quantum simulation workflows.
The proposed framework employs an LSTM network followed by a fully connected (FC) projection layer that produces ansatz parameters in a size-adaptive manner, enabling a single meta-learner to generalize across molecular systems of varying complexity.

All chemistry benchmarks are executed on NVIDIA H100 GPUs, while physics-based Simple Harmonic Motion (SHM) experiments are performed on an NVIDIA RTX~5090 GPU, reflecting the differing computational demands of the two problem classes.

By deploying this pipeline, we demonstrate near Full Configuration Interaction (FCI) accuracy for molecular ground states while maintaining favorable scaling with system size.
Beyond chemistry, we apply the same framework to physics-based SHM systems, including the Simple Harmonic Oscillator (SHO), where exact solutions are known.
Using VQE and VQD, the model successfully reproduces both ground and excited states, providing a controlled testbed for analyzing optimization dynamics.
Together, these results highlight how GPU acceleration and meta-learning stabilize variational optimization and reduce quantum resource requirements.

In summary, this work makes three primary contributions. 
First, we introduce a size adaptive LSTM-FC meta initialization strategy that generalizes across molecular systems with varying ansatz dimensionality, priors across problem sizes. 
Second, we integrate the proposed VQE-LSTM framework with NVIDIA’s CUDAQ backend, achieving full GPU acceleration of Hamiltonian evaluation and classical optimization within the variational loop. 
Third, we validate the framework on both chemical and physics-based Hamiltonians, demonstrating accurate ground- and excited-state estimation with reduced optimization cost and improved convergence reliability.

\section{Methodology}

\subsection{Variational Quantum Eigensolver (VQE)}

The VQE is a hybrid quantum--classical algorithm for approximating extremal eigenvalues of a Hamiltonian using a parameterized quantum circuit optimized by a classical routine.
In this work, we apply VQE to two distinct problem classes: molecular electronic-structure Hamiltonians from quantum chemistry and physics-based many-body Hamiltonians derived from the simple harmonic oscillator (SHO).
This dual setting allows us to evaluate both physical accuracy and cross-domain generalization of the proposed meta-initialization strategy.

Given a Hamiltonian $\hat{H}$ and a parameterized ansatz $U(\boldsymbol{\theta})$, VQE minimizes the energy expectation value
\begin{equation}
E(\boldsymbol{\theta})=\langle 0 | U^{\dagger}(\boldsymbol{\theta}) \hat{H} U(\boldsymbol{\theta}) | 0 \rangle
\label{eq:vqe_energy}
\end{equation}
using a classical optimizer. All expectation values and gradients are evaluated using GPU accelerated simulation provided by CUDAQ.

\subsubsection{Hamiltonians}

\paragraph{Chemistry Hamiltonians}
Molecular Hamiltonians are generated using \texttt{PySCF}, which computes one- and two-electron integrals defining the second-quantized fermionic Hamiltonian
\begin{equation}
\hat{H}_{\mathrm{chem}} =
\sum_{p,q} h_{pq}\, \hat{a}^{\dagger}_{p}\hat{a}_{q}
+
\frac{1}{2}\sum_{p,q,r,s}
h_{pqrs}\, \hat{a}^{\dagger}_{p}\hat{a}^{\dagger}_{q}
\hat{a}_{r}\hat{a}_{s}.
\label{eq:chem_ham}
\end{equation}
CUDAQ performs fermion-to-qubit mapping (Jordan--Wigner or Bravyi--Kitaev), producing a qubit Hamiltonian expressed as a weighted sum of Pauli operators,
\begin{equation}
\hat{H}_{\mathrm{chem}} = \sum_j w_j P_j,
\label{eq:chem_pauli}
\end{equation}
which is directly used in the VQE optimization loop.
Chemistry results are benchmarked against FCI when feasible, and otherwise against CASCI references.
\paragraph{Physics Hamiltonians}
To evaluate performance beyond chemistry, we apply VQE to a physics-motivated SHO benchmark.
The continuous-variable Hamiltonian is
\begin{equation}
\hat{H}_{\mathrm{phys}} =
\frac{\hat{p}^{2}}{2m}
+
\frac{1}{2} m \omega^{2} \hat{x}^{2}
=
\omega\left(\hat{a}^{\dagger}\hat{a} + \tfrac{1}{2}\right),
\label{eq:sho_ham}
\end{equation}
where $\hat{a}$ and $\hat{a}^{\dagger}$ are ladder operators satisfying
\begin{equation}
\hat{a}\ket{n} = \sqrt{n}\ket{n-1},
\qquad
\hat{a}^{\dagger}\ket{n} = \sqrt{n+1}\ket{n+1}.
\label{eq:ladder_ops}
\end{equation}

We construct a finite-dimensional matrix representation of $\hat{H}_{\mathrm{phys}}$ in an energy-level basis and embed it into an $n$-qubit Hilbert space of dimension $2^n$.
Thus, computational basis states index oscillator energy levels rather than spatial degrees of freedom, and increasing $n$ enlarges the accessible Hilbert space without changing the operator form of $\hat{H}_{\mathrm{phys}}$.
The embedded operator is then converted into a weighted Pauli expansion,
\[
\hat{H}_{\mathrm{phys}} = \sum_k \alpha_k P_k,
\]
and evaluated using the same CUDAQ VQE pipeline as for chemistry systems.

\paragraph{Variational Quantum Deflation (VQD)}
Excited states are computed using VQD, which augments the VQE objective with an overlap penalty.
For the first excited state, the objective is
\begin{equation}
C_{\mathrm{VQD}}(\boldsymbol{\theta})
=
\langle \psi(\boldsymbol{\theta}) | \hat{H} | \psi(\boldsymbol{\theta}) \rangle
+
\beta \left| \langle \psi(\boldsymbol{\theta}) | \psi_{0} \rangle \right|^{2},
\label{eq:vqd_obj}
\end{equation}
where $\psi_{0}$ is the previously optimized ground state.
Minimizing this objective yields an approximate excited eigenstate orthogonal to $\psi_{0}$.
\subsubsection{Ansatz Structures}

\paragraph{Hardware-Efficient Ansatz (HEA)}
For physics-based Simple Harmonic Motion (SHM) experiments, we employ a hardware-efficient ansatz consisting of repeated layers of single-qubit rotations and entangling gates.
Each layer applies $R_y$ and $R_z$ rotations to every qubit, followed by a fully connected entangler:
\begin{equation}
U_{\mathrm{HEA}}(\boldsymbol{\theta})
=
\prod_{l=1}^{L}
\left[
\prod_{i=1}^{n}
R_y(\theta^{(1)}_{l,i})
R_z(\theta^{(2)}_{l,i})
\;
\prod_{i<j}
\mathrm{CNOT}(i,j)
\right].
\label{eq:hea}
\end{equation}
The total number of parameters scales linearly with the number of qubits and layers.
The HEA is additionally used in selected chemistry experiments to evaluate ansatz-dependent optimization behavior.

\paragraph{Unitary Coupled Cluster with Singles and Doubles (UCCSD)}
All quantum chemistry benchmarks employ the UCCSD ansatz, which is derived from electronic-structure theory and systematically captures electron correlation.
The ansatz is defined as
\begin{equation}
U_{\mathrm{UCCSD}}(\boldsymbol{\theta})
=
\exp\!\left(
\hat{T}_{1}(\boldsymbol{\theta})
+
\hat{T}_{2}(\boldsymbol{\theta})
-
\mathrm{h.c.}
\right),
\label{eq:uccsd}
\end{equation}
where $\hat{T}_{1}$ and $\hat{T}_{2}$ generate single and double fermionic excitations, respectively.
After fermion-to-qubit mapping, the resulting excitation operators are implemented using exponentiated Pauli strings within CUDAQ.
UCCSD is not applied to the SHM system.

\subsection{LSTM-FC Meta-Initialization}

To reduce sensitivity to poor initial parameters, we introduce an LSTM-FC meta-learner that predicts VQE initialization parameters.
The LSTM processes sequences of previously observed energies and parameter updates, while a final fully connected layer outputs a fixed-length parameter vector.

To support systems with varying ansatz sizes, we fix the parameter dimension to that of the largest system considered.
Smaller systems are zero-padded during training, and only the required prefix of the predicted vector is used during inference.

This padding-and-slicing strategy enables a single LSTM-FC model to generalize across different system sizes without retraining or molecule-specific heads, yielding transferable optimization priors that scale naturally with problem size~\cite{chang2025lstmfcvqe}.

\begin{figure*}[t]
\centering
\includegraphics[width=\textwidth]{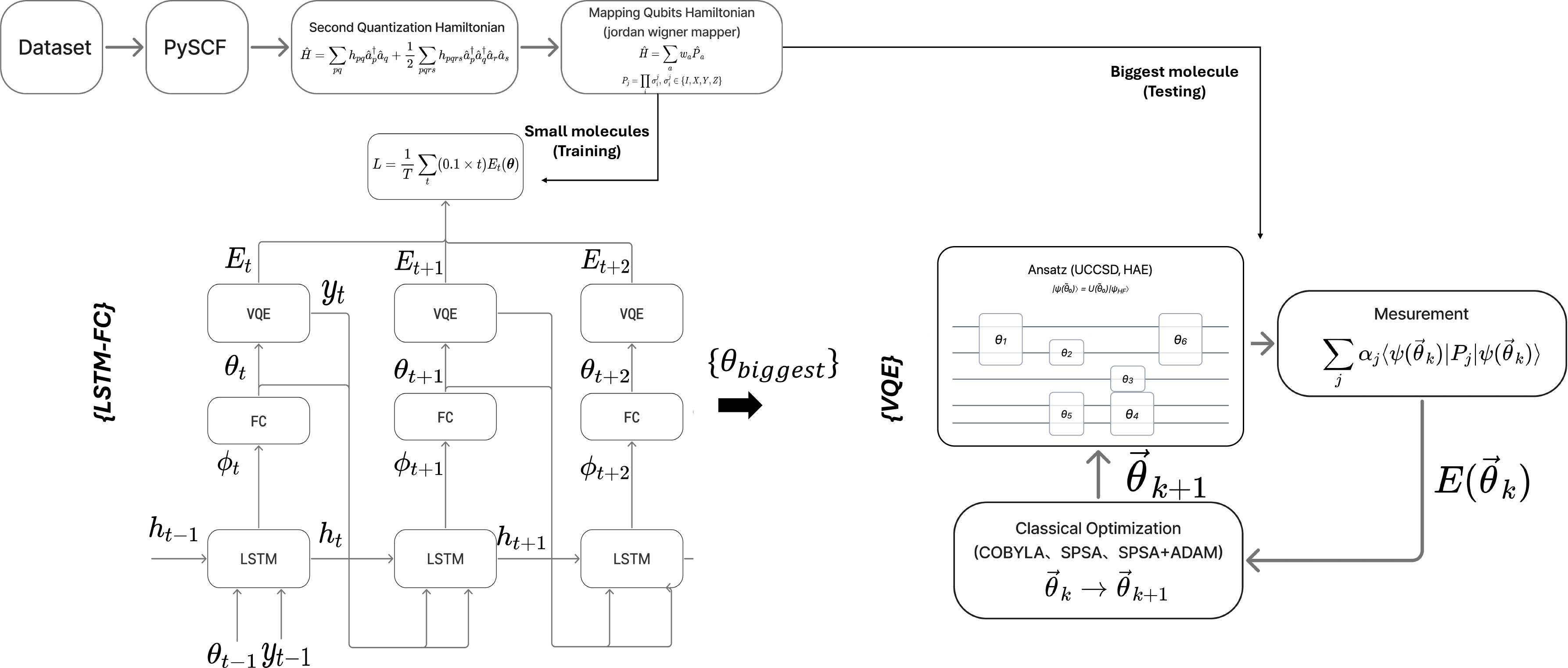}
\caption{
Workflow of the proposed LSTM-VQE framework. During meta-training, the framework module learns a mapping from prior optimization traces with each FC iteration. At inference, this predicted initialization is passed to a standard VQE, where CUDAQ evaluates expectation values on GPU and a classical optimizer refines parameters to convergence.
}
\label{fig:flowchart}
\end{figure*}

\subsection{Framework}
Figure~\ref{fig:flowchart} illustrates the complete LSTM–VQE workflow. 
The process begins with Hamiltonian construction using \texttt{PySCF}, followed by qubit mapping and GPU-based simulation via CUDAQ. 
The LSTM–FC module generates informed initial parameters, which are then refined through classical optimization within the VQE loop. 
Final energies are compared against reference solutions such as FCI or CASCI.

\section{Results}

\subsection{Quantum Chemistry Benchmarks}
We evaluate the proposed LSTM-FC initialization strategy on molecular Hamiltonians generated by \texttt{PySCF}. Unless otherwise noted, chemistry benchmarks use UCCSD; HEA results are included only to assess ansatz dependent optimization behavior. All chemistry results report energy deviation relative to a classical reference, along with convergence runtime under CPU and GPU backends.


\begin{figure}[h!]
\centering
\includegraphics[width=0.9\columnwidth]{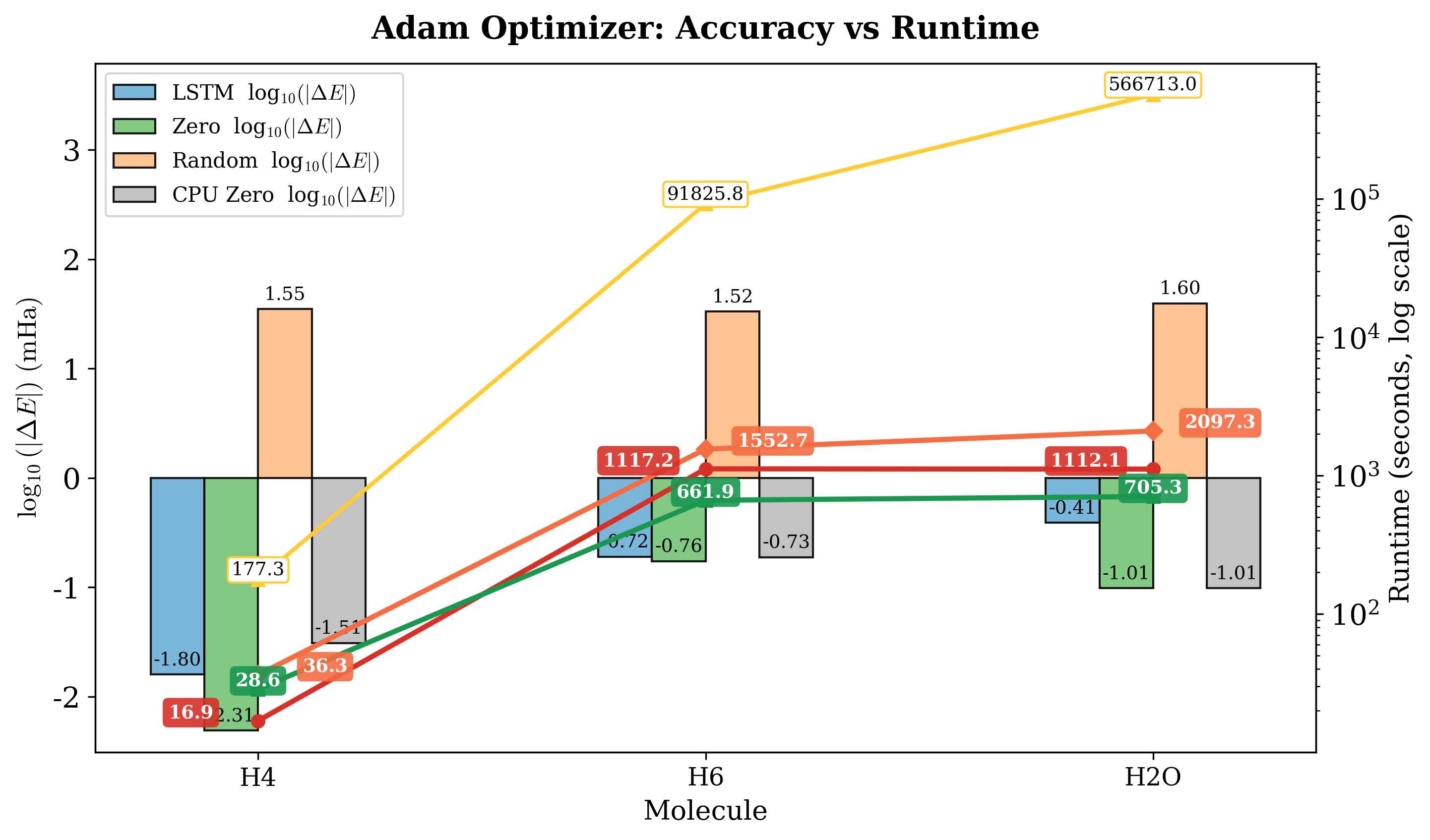}
\includegraphics[width=0.9\columnwidth]{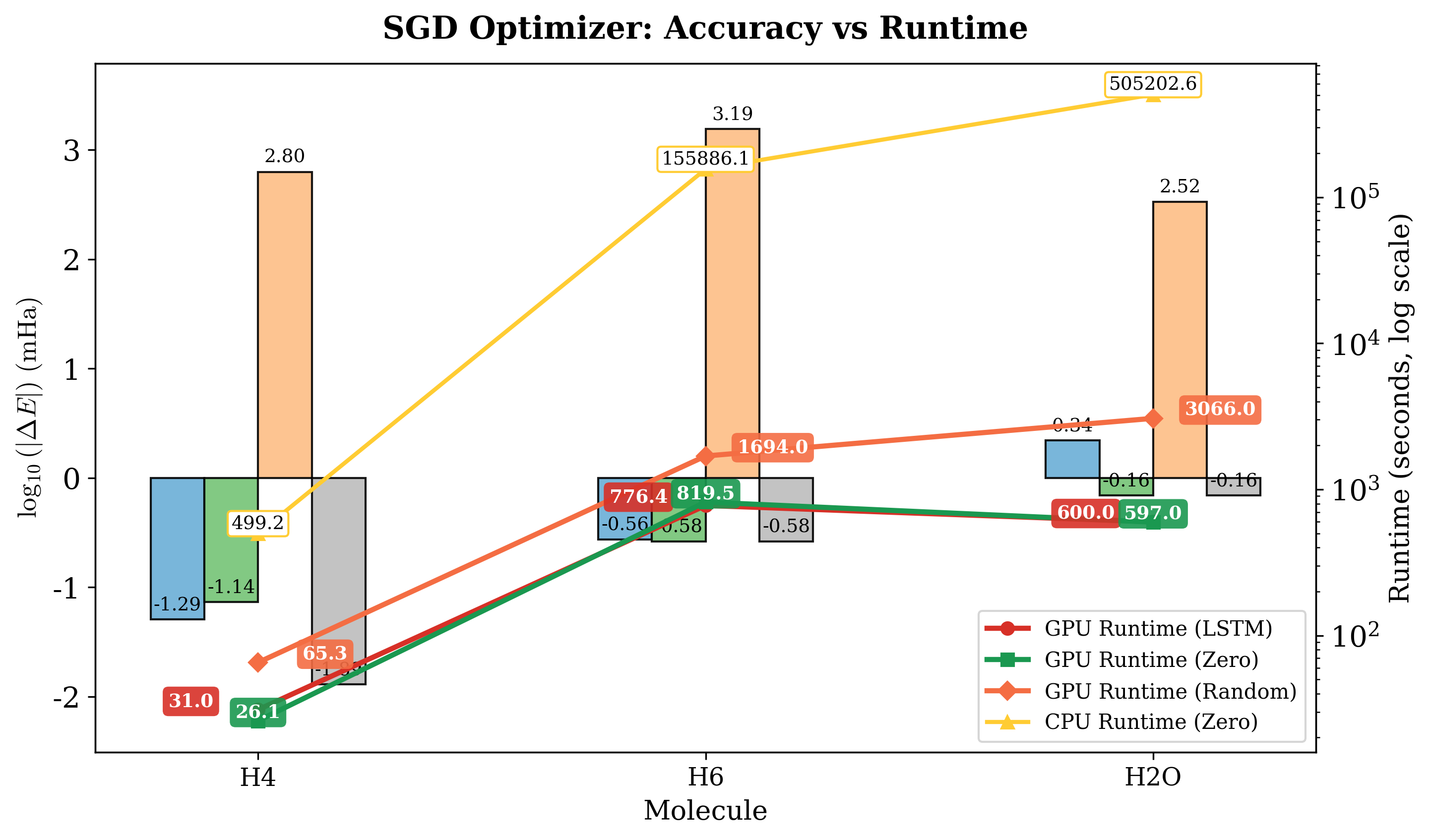}
\caption{
Energy deviation and runtime comparison under Adam (top) and SGD (bottom) optimizers. Bars show $\log_{10}(|\Delta E|)$ relative to the classical reference; lines indicate runtime for GPU and CPU backends. LSTM initialization improves convergence reliability and reduces runtime sensitivity across molecules.
}
\label{fig:optimizer_comparison}
\end{figure}
Figure~\ref{fig:optimizer_comparison} compares energy deviation and runtime for Adam and SGD under LSTM-based, all-zero, and random initialization strategies. Across all molecules, LSTM initialization improves or matches the best accuracy and reduces runtime variance relative to both random and all-zero initialization. Adam is used as the primary optimizer in subsequent experiments because it converges more consistently on average across molecules and trials. GPU acceleration yields over an $800\times$ speedup for H$_2$O under zero initialization and becomes increasingly critical for larger systems, where CPU-based optimization is often impractical.

A key observation is that the benefits of LSTM initialization are not tied to a single optimizer. 
Both Adam and SGD exhibit improved stability when warm-started by the meta-learned parameters. 
Meanwhile, the GPU backend significantly reduces total runtime in all settings due to accelerated expectation evaluation, making repeated circuit evaluations feasible even at larger qubit counts.


\begin{figure}[h!]
\centering
\includegraphics[width=0.9\columnwidth]{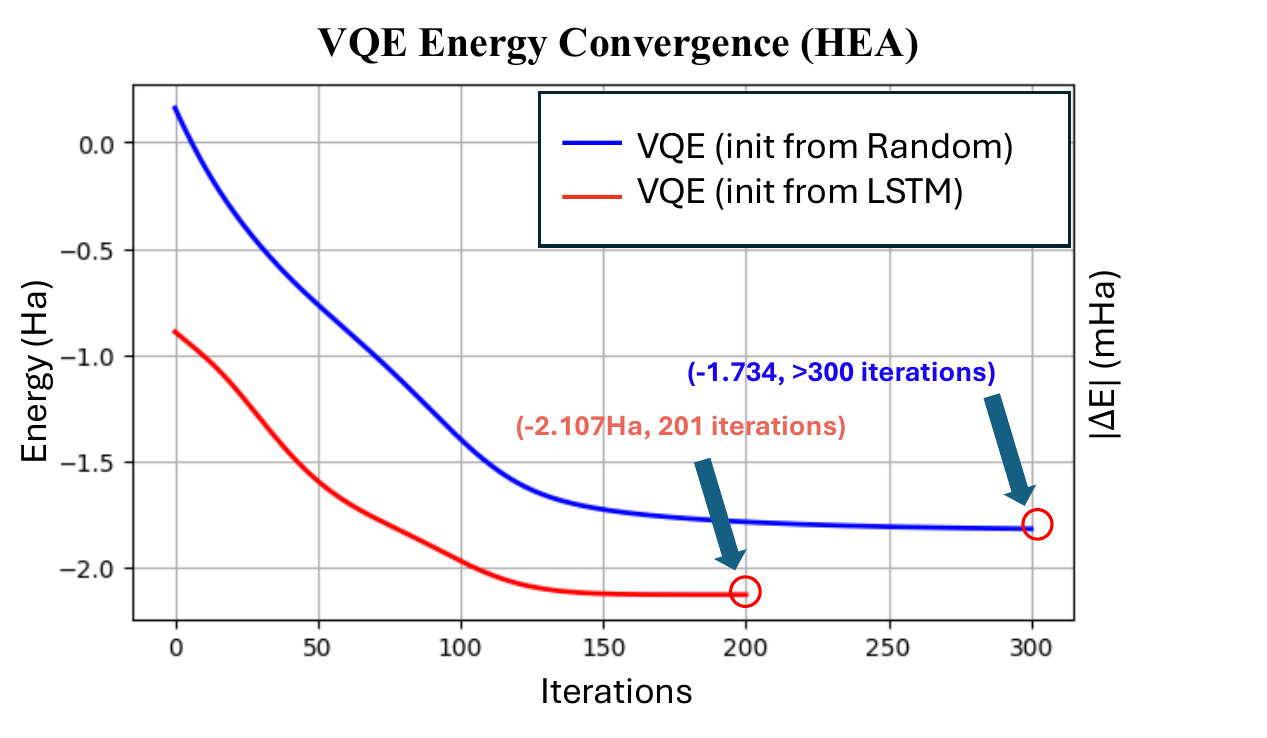}
\caption{
Representative VQE convergence for H$_4$ under HEA. LSTM initialization accelerates entry into a low-energy basin and reduces the number of iterations required to reach a stable solution compared to conventional initialization.
}
\label{fig:h4_convergence}
\end{figure}

To illustrate how initialization affects the optimization trajectory, Fig.~\ref{fig:h4_convergence} shows a representative convergence curve (H$_4$ under the HEA ansatz). 
The LSTM-initialized run reaches a low-energy basin substantially earlier than random initialization and avoids long plateaus.


\begin{figure}[h!]
\centering
\includegraphics[width=0.85\columnwidth]{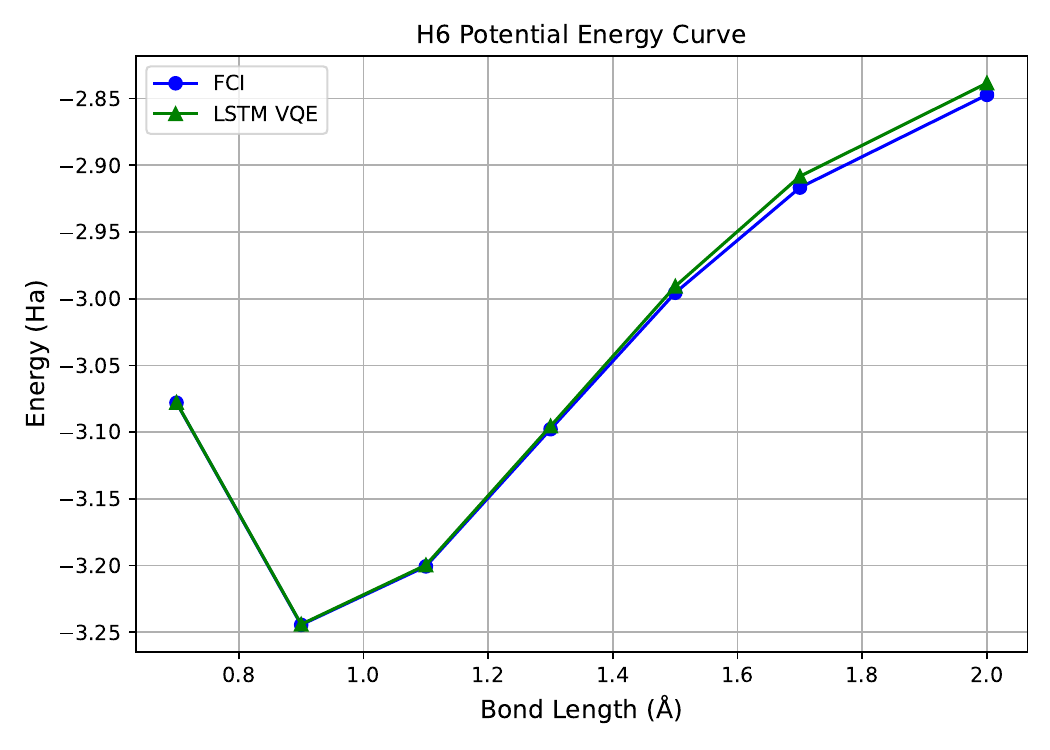}
\caption{
H$_6$ potential energy curve (PEC) as a function of bond length. LSTM-initialized VQE reproduces the reference curve across the scan range, demonstrating robustness under continuous Hamiltonian variation.
}
\label{fig:h6_pec}
\end{figure}

Beyond single-geometry benchmarks, a practical electronic-structure solver must remain stable under continuous Hamiltonian variation, as encountered in geometry scans and reaction pathways. 
Figure~\ref{fig:h6_pec} reports the H$_6$ potential energy curve (PEC) across multiple bond lengths. 
Across the full scan range, LSTM--VQE closely tracks the classical reference, indicating that the learned initialization generalizes beyond a fixed molecular geometry and remains consistent as the electronic structure changes smoothly.

This result is important for repeated-evaluation workloads (e.g., geometry optimization), where each Hamiltonian differs slightly from the previous one In such settings, improved initialization can compound into substantial total runtime reduction.


\begin{figure}[h!]
\centering
\includegraphics[width=0.9\columnwidth]{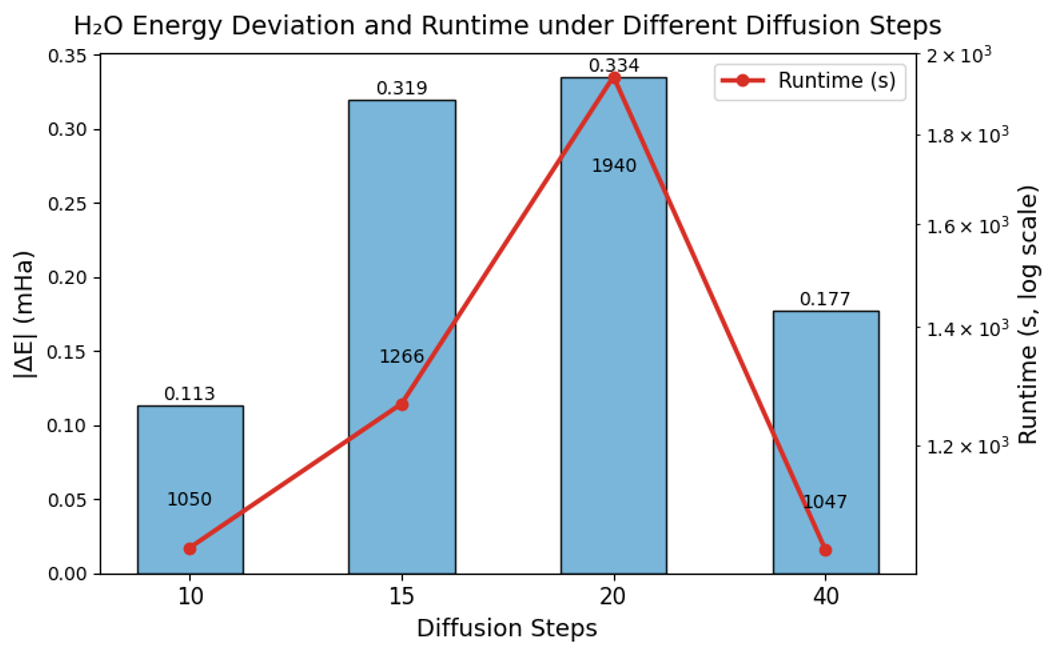}
\caption{
Impact of diffusion steps on H$_2$O performance. Bars show $|\Delta E|$ and the line shows runtime. Larger diffusion depth increases overhead and does not guarantee improved accuracy.
}
\label{fig:h2o_diffusion}
\end{figure}

In our implementation, diffusion steps correspond to the number of recurrent unroll steps through the LSTM--FC module prior to launching the VQE optimizer. 
Figure~\ref{fig:h2o_diffusion} shows the resulting trade-off between accuracy and runtime for H$_2$O. 
Increasing the number of diffusion steps does not monotonically improve energy accuracy; instead, moderate diffusion can be sufficient, while deeper diffusion introduces additional overhead and may over-adjust the parameter guess.

This suggests that the meta-initialization depth should be treated as a tunable hyperparameter, and that the model benefits more from producing a good basin entry point than from repeatedly refining the guess without feedback from the quantum objective.


\begin{figure}[h!]
\centering
\includegraphics[width=0.9\columnwidth]{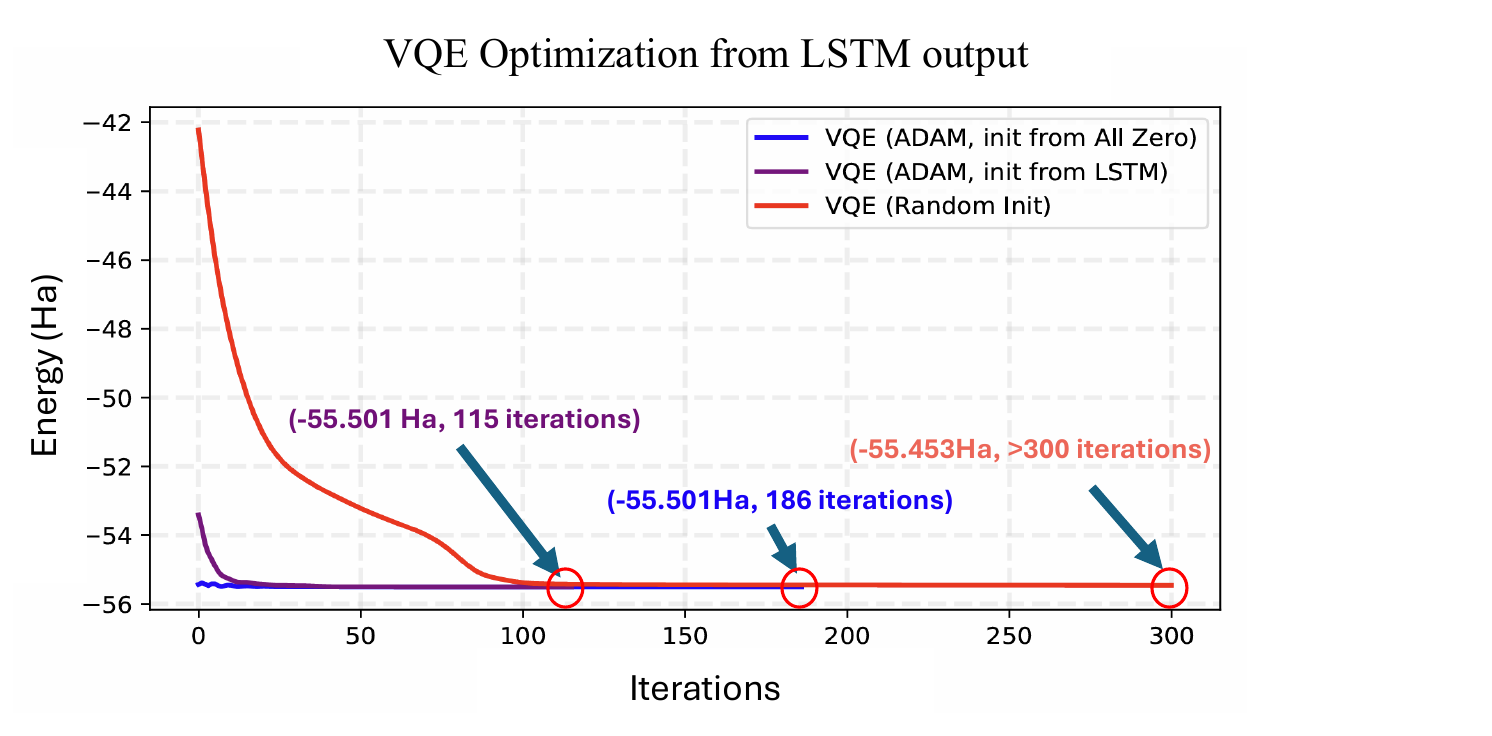}
\caption{
VQE convergence for 16-qubit NH$_3$. LSTM initialization converges within the iteration budget, while random initialization fails to achieve comparable convergence within the same limit.
}
\label{fig:nh3_convergence}
\end{figure}

We next evaluate out-of-distribution scaling by testing a 16-qubit NH$_3$ instance while training only on smaller molecules (H$_2$, H$_3$, H$_4$, H$_6$, OH$^-$). 
Figure~\ref{fig:nh3_convergence} shows that random initialization fails to reach a comparable energy within the iteration budget, whereas LSTM initialization achieves the best convergence along with the fastest observed GPU runtime (6228.4 s). 
These results highlight the combined benefit of meta-learned initialization and GPU acceleration, which becomes increasingly critical as system size grows and CPU-based optimization becomes impractical.

\begin{figure}[h!]
\centering
\includegraphics[width=0.9\columnwidth]{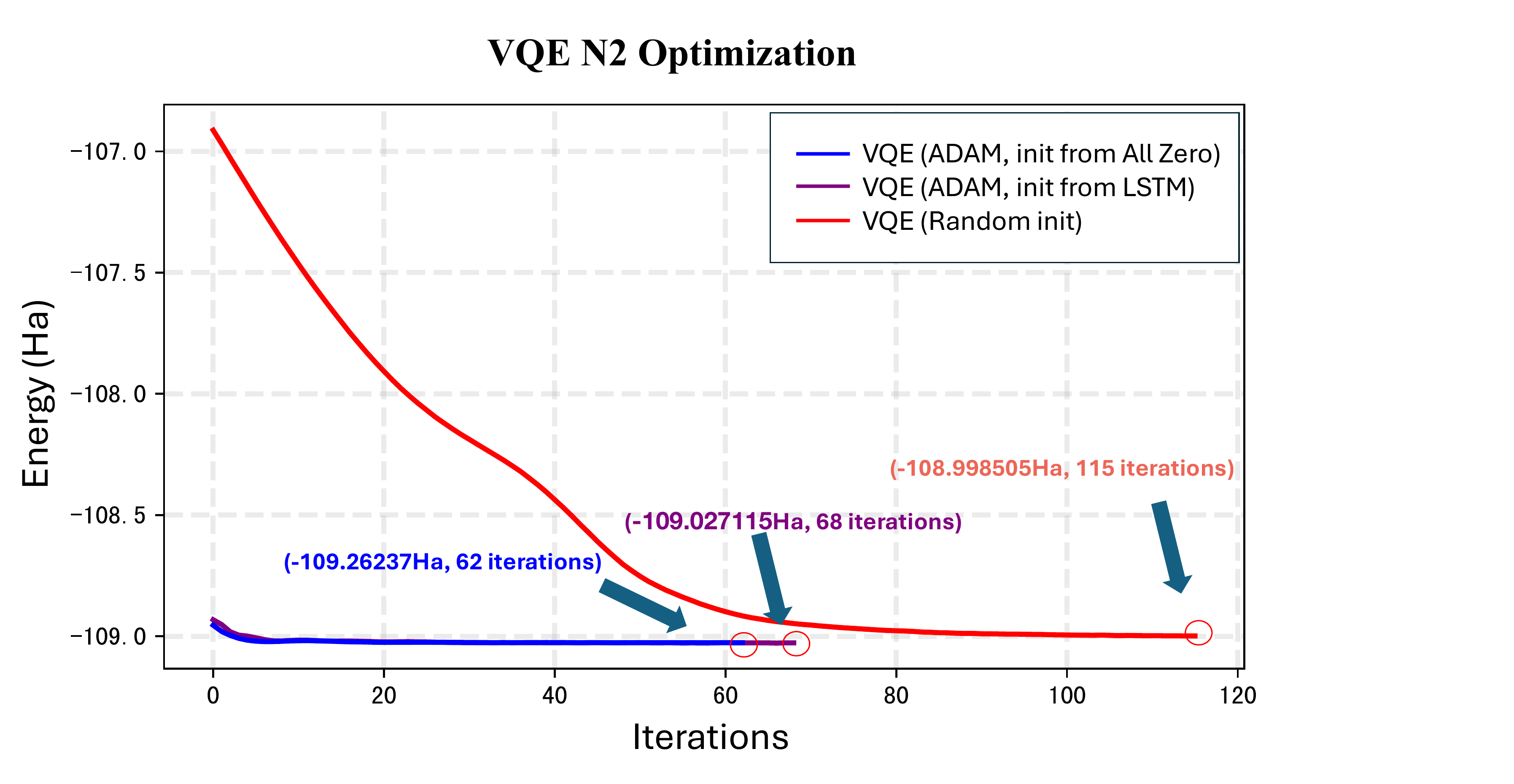}
\caption{
VQE convergence for N$_2$ in an active space (CAS 10e/7o, 14 qubits, cc-pVDZ). LSTM initialization reaches the lowest final energy among tested initializations and converges more reliably than random initialization.
}
\label{fig:n2_convergence}
\end{figure}

To evaluate chemically realistic scalability beyond small full-space cases, we consider N$_2$ using an active-space reducting (CAS 10e/7o in cc-pVDZ), from a 20 to 14-qubit Hamiltonian. 
We use FCI when feasible, but here we use CASCI within active space for larger system references.
Figure~\ref{fig:n2_convergence} reports convergence behavior under different initializations. 
LSTM initialization converges to $-109.0271$ Ha, within $4.56\times10^{-4}$ Ha of the CASCI reference ($-109.0275$ Ha), while also reducing iteration count and runtime relative to random initialization. Despite having a higher initialization relative to all-zero, it resulted in a more accurate convergence.

Overall, this experiment indicates that the LSTM-VQE pipeline remains effective when paired with standard quantum chemistry approximations such as active-space selection, which is essential for pushing toward larger systems.

\begin{table}[h!]
\centering
\scriptsize
\caption{VQE Learning Rate (LR) Scan Summary for SHO System}
\label{tab:vqe_lr_scan}
\setlength{\tabcolsep}{3pt} 
\renewcommand{\arraystretch}{1.1}
\begin{tabular}{cccccc}
\hline
\textbf{LR} & \textbf{Final $E_{\mathrm{VQE}}$ (a.u.)} & \textbf{$E_{\mathrm{exact}}$ (a.u.)} & \textbf{Abs. Error (a.u.)} & \textbf{Iter.} & \textbf{Time (s)} \\
\hline
$1\times10^{-3}$ & 0.25001 & 0.25000 & $5.84\times10^{-6}$ & 21  & 0.37 \\
$3\times10^{-4}$ & 0.25000 & 0.25000 & $1.28\times10^{-6}$ & 74  & 1.31 \\
$1\times10^{-4}$ & 0.25001 & 0.25000 & $7.48\times10^{-6}$ & 159 & 2.86 \\
$3\times10^{-5}$ & 0.25003 & 0.25000 & $3.17\times10^{-5}$ & 324 & 5.87 \\
$1\times10^{-5}$ & 0.25008 & 0.25000 & $7.60\times10^{-5}$ & 545 & 9.91 \\
$3\times10^{-6}$ & 0.25026 & 0.25000 & $2.62\times10^{-4}$ & 89  & 1.58 \\
$1\times10^{-6}$ & 0.25027 & 0.25000 & $2.74\times10^{-4}$ & 1   & 0.02 \\
\hline
\multicolumn{6}{l}{\textbf{System:} SHO \quad \textbf{n\_qubits:} 4 \quad \textbf{Fixed Param Dim:} 40 \quad \textbf{Tol:} $1\times10^{-7}$} \\
\multicolumn{6}{l}{\textbf{Best LR:} $3\times10^{-4}$ \quad \textbf{Best Abs. Err:} $1.28\times10^{-6}$} \\
\hline
\end{tabular}
\end{table}

\subsection{Simple Harmonic Motion (SHM)}

We further evaluate the framework on a physics motivated SHO benchmark, where analytical energies provide a clean reference. 
Unlike chemistry Hamiltonians, the SHO setting isolates optimization and expressivity effects without electronic-structure complexity, enabling controlled study of convergence, learning-rate sensitivity, and the extension to excited states via VQD. 
In all SHM experiments, we employ HEA ansatz to parameterize the variational circuit.


Table~\ref{tab:vqe_lr_scan} reports a learning-rate scan for the classical optimizer under a SHO Hamiltonian ($n=4$ qubits). 
An intermediate learning rate achieves the best balance between convergence speed and accuracy and is used for subsequent SHM experiments.


\begin{figure}[h!]
\centering
\includegraphics[width=0.9\columnwidth]{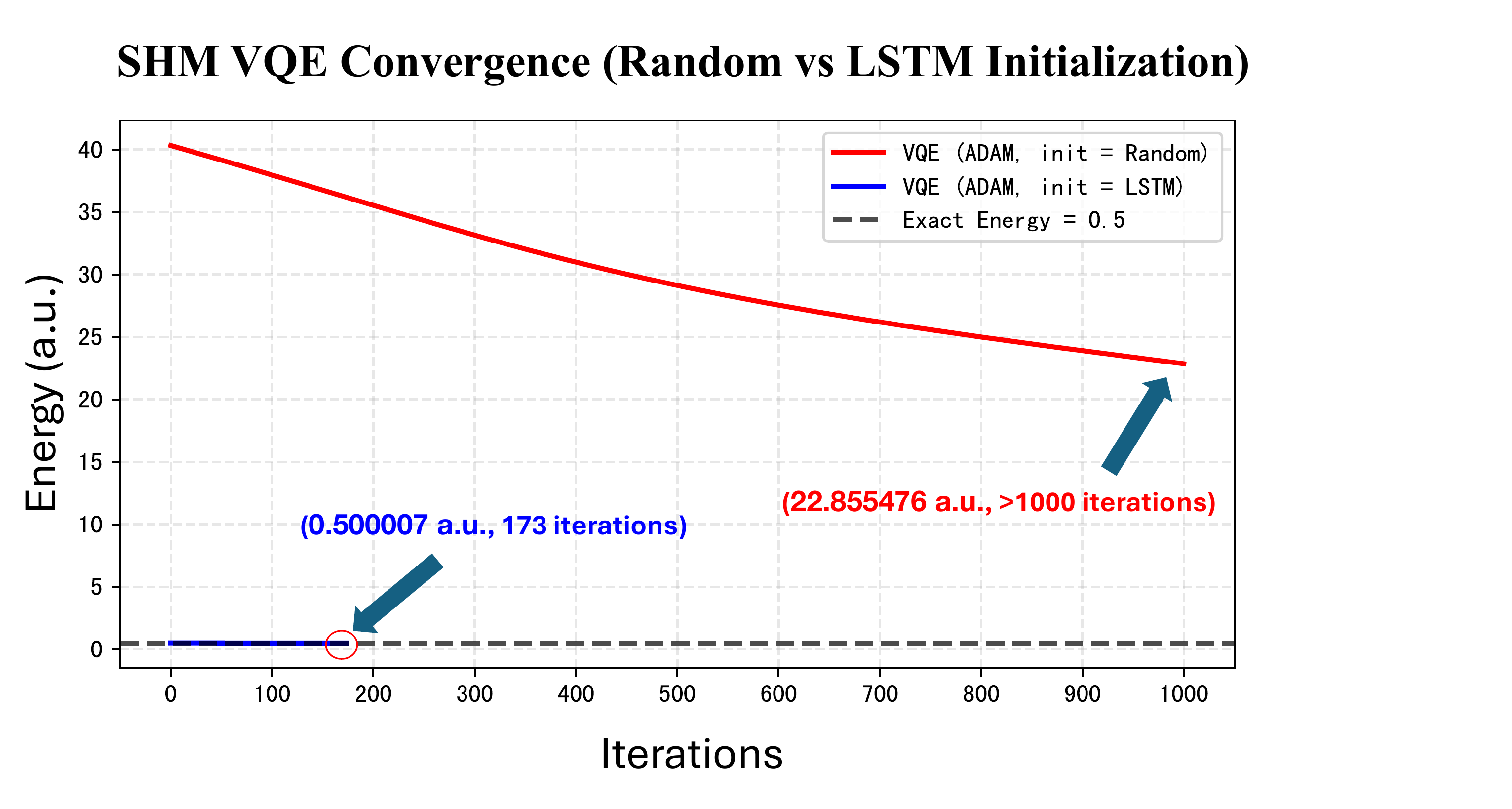}
\caption{
Ground-state VQE convergence for SHO under random and LSTM initialization. LSTM initialization reaches the target energy basin with fewer iterations and improved stability.
}
\label{fig:vqe_SHM}
\end{figure}

Figure~\ref{fig:vqe_SHM} compares convergence trajectories under random and LSTM initialization. 
The LSTM-initialized run rapidly converges toward the correct ground-state basin, while random initialization requires substantially more iterations and exhibits larger fluctuations. 
This demonstrates that the meta-initialization mechanism transfers beyond chemistry-specific structure and can accelerate convergence in a distinct physics Hamiltonian family.

\begin{table}[h!]
\centering
\scriptsize
\caption{Runtime Comparison between GPU (CUDAQ) and CPU Execution for SHO ($n_\text{qubits}=6$)}
\label{tab:gpu_cpu_runtime}
\setlength{\tabcolsep}{2pt} 
\renewcommand{\arraystretch}{1.1}
\begin{tabular}{lcccc}
\hline
\textbf{Initialization} & \textbf{Iterations} & \textbf{GPU Runtime (s)} & \textbf{Est. CPU Runtime (s)} \\
\hline
LSTM Init   & 173  & 25.14  & \textbf{12,205.0} \\
Random Init & 1000 & 144.02  & \textbf{69,070.0} \\
\hline
\multicolumn{4}{l}{\textbf{Speedup (GPU vs. CPU):} $\sim$486$\times$ (LSTM), $\sim$479$\times$ (Random)} \\
\hline
\end{tabular}
\end{table}


Table~\ref{tab:gpu_cpu_runtime} summarizes runtime and iteration counts for the SHO benchmark on GPU (CUDAQ) and CPU backends. 
The GPU backend provides substantial acceleration due to high-throughput expectation evaluation, and the reduced iteration count from LSTM initialization compounds this benefit. 
Together, these effects enable practical scaling to more expensive variational workloads.


\begin{figure}[h!]
\centering
\includegraphics[width=0.95\columnwidth]{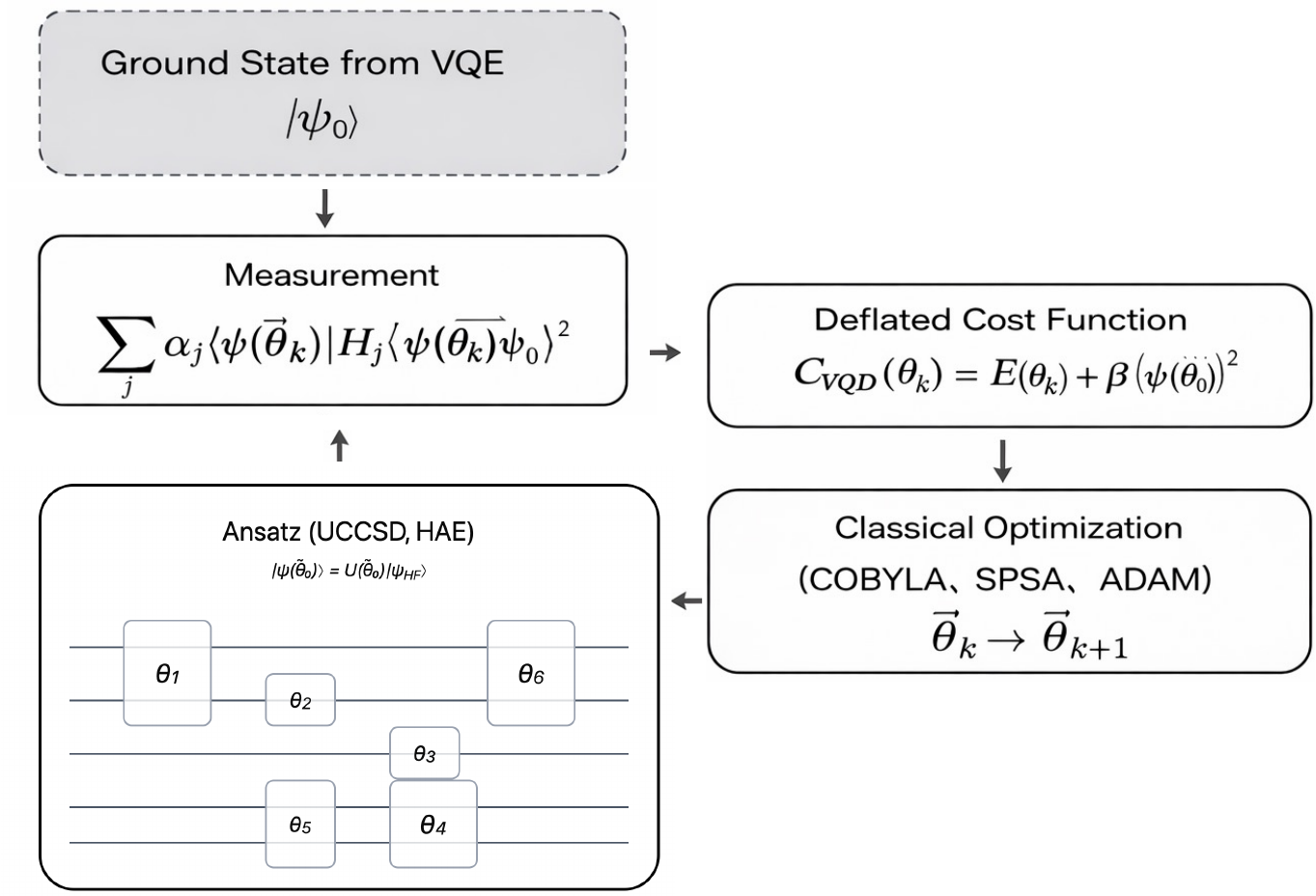}
\caption{
VQD workflow following VQE. The ground state $\ket{\psi_0}$ is first obtained via VQE.
VQD then minimizes a deflated objective consisting of the Hamiltonian expectation plus an overlap penalty
to suppress convergence back to $\ket{\psi_0}$.
}
\label{fig:vqd_flowchart}
\end{figure}

Figure~\ref{fig:vqd_flowchart} illustrates the VQD pipeline used in this work.
We first obtain the ground state via VQE to construct a reference state $\ket{\psi_0}$.
The VQD objective then augments the energy expectation with a deflation penalty
$\beta |\langle \psi(\boldsymbol{\theta}) \mid \psi_0 \rangle|^2$,
encouraging the optimizer to converge toward an excited state while suppressing overlap with the ground state.
The resulting optimized parameters are used to estimate the excited-state energy.

\begin{figure}[h!]
    \centering
    \includegraphics[width=\linewidth]{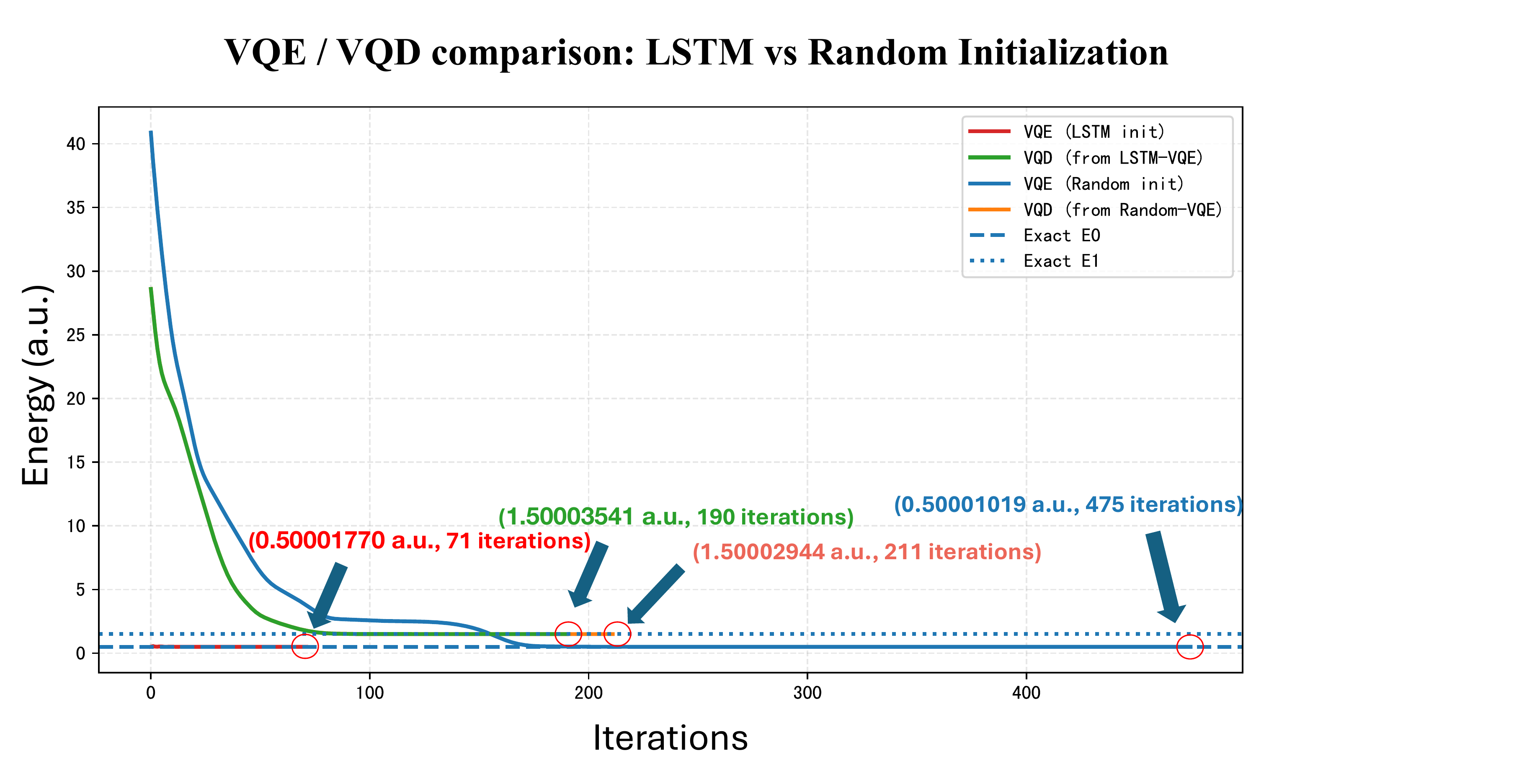}
    \caption{
    Excited-state optimization of the Simple Harmonic Oscillator using VQD.
    Both random and LSTM-initialized runs converge to the first excited-state energy,
    with LSTM initialization exhibiting faster and more stable convergence.
    }
    \label{fig:vqd_SSS}
\end{figure}

We extend the framework to excited-state computation using VQD.
Starting from the VQE optimized ground state, VQD augments the objective function with an overlap penalty to prevent reconvergence to lower-energy states.
As shown in Fig.~\ref{fig:vqd_SSS}, the optimization converges to the first excited-state energy under both random and meta-learned initialization.

At convergence, the squared overlap between the optimized excited-state wavefunction and the ground state is strongly suppressed,
with final values of $2.84\times10^{-8}$ for LSTM initialization and $9.61\times10^{-9}$ for random initialization.
These results confirm that the deflation term effectively enforces orthogonality, enabling reliable excited-state targeting within the proposed meta-learning framework.

\section{Conclusion}
We presented a GPU accelerated VQE–LSTM framework that integrates meta-learned initialization with NVIDIA’s CUDAQ backend to improve the scalability of variational quantum simulation across chemistry and physics domains. Across multiple systems, LSTM–FC initialization converges faster and more reliably than standard initializations. Stronger ground-state convergence further stabilizes VQD for excited-state optimization, and GPU accelerated expectation evaluation significantly reduces runtime, enabling exploration of deeper circuits and larger qubit counts. Together, these results demonstrate that combining learned priors with high-performance simulation infrastructure provides a practical path toward scalable variational algorithms.
Accurate molecular energies support applications in materials discovery and drug design, while SHM benchmarks provide a controlled setting to analyze optimization behavior for general quantized Hamiltonians. Overall, the proposed pipeline shows how learned initialization and GPU acceleration jointly enhance convergence reliability, runtime efficiency, and scalability for many-body quantum simulations.
As future work, we plan to extend the framework to more complex systems, including quantum pendulum and lattice models, improve multi-excited-state VQD, and investigate scaling under larger GPU resources such as multi-GPU acceleration.

\bibliographystyle{IEEEtran}
\bibliography{vqe_lstm_refs}

\end{document}